\documentclass[epsfig,floats,pre,twocolumn,superscriptaddress,
preprintnumbers,floatfix]{revtex4-1}

\usepackage[latin2]{inputenc}
\usepackage{amsmath}
\usepackage{amssymb}
\usepackage{epsfig}
\usepackage{bm}
\usepackage{color}
\usepackage{stackengine}
\usepackage{graphicx}
\usepackage{caption, subcaption}
\usepackage{sidecap}

\begin{document}

\title{\textbf{Vulnerability of Transport through Evolving Spatial Networks}}

\author{Ali Molavi}
\affiliation{Department of Physics, K.N.\ Toosi University of Technology, 
Tehran 15875-4416, Iran}
\author{Hossein Hamzehpour}
\affiliation{Department of Physics, K.N.\ Toosi University of Technology, 
Tehran 15875-4416, Iran}
\author{Reza Shaebani}
\affiliation{Department of Theoretical Physics and Center for Biophysics, 
Saarland University, 66123 Saarbr\"ucken, Germany}

\begin{abstract}\noindent
Insight into the blockage vulnerability of evolving spatial networks is important 
for understanding transport resilience, robustness, and failure of a broad class 
of real-world structures such as porous media and utility, urban traffic, and 
infrastructure networks. By exhaustive search for central transport hubs on 
porous lattice structures, we recursively determine and block the emerging main 
hub until the evolving network reaches the impenetrability limit. We find that 
the blockage backbone is a self-similar path with a fractal dimension which is 
distinctly smaller than that of the universality class of optimal path crack 
models. The number of blocking steps versus the rescaled initial occupation 
fraction collapses onto a master curve for different network sizes, allowing 
for the prediction of the onset of impenetrability. The shortest-path length 
distribution broadens during the blocking process reflecting an increase of 
spatial correlations. We address the reliability of our predictions upon 
increasing the disorder or decreasing the fraction of processed structural 
information. 
\end{abstract}
 
\maketitle

\section{Introduction}

Spatial networks represent systems in which the interconnected structure is embedded 
in real space. Examples range from transportation networks, power grids, and pedestrian 
crowds to granular materials and porous structures \cite{Barthelemy11}. Spreading 
and transport processes on spatial networks have attracted much attention from both 
scientific and technological points of view \cite{Colak16,Ren14,Morris12,Gonzalez08,
Shaebani14,KartunGiles19,Kang11,Zhang19,Li15,Shaebani18,Carmona20,Zeng19,Olmos18,Fouladvand04}. 
Particularly, material transport through porous structures has been one of the basic 
problems of statistical physics \cite{Hunt17,Meakin09}. Despite the simple topology 
of spatial networks (compared to complex networks), transport on such structures 
exhibits an intriguing complexity due to the relevance of spatial dimensions, 
correlations and constraints, and dynamical evolution of structures. Achieving 
an optimal transport through a given spatial network requires a computationally 
costly analysis of structural information. Despite efforts (such as multiscale 
network analysis approach \cite{ErcseyRavasz10}), it is still an open challenge 
to reduce the huge computational costs by, for example, resorting to a partial 
information processing or taking advantage of structural self-similarities.

Optimal path models \cite{Hansen04,Andrade09,Oliveira11,Fehr12,SampaioFilho18,Talon13,Folz23}--- 
which minimize a cost function along possible routes in a disordered cost landscape--- 
have been broadly employed in technological applications, e.g.\ internet routing and 
urban traffic, or to mimic natural processes such as transport in porous media, 
fracturing of heterogeneous materials, or electrical current through random structures 
\cite{Carmona20,Moreira12,Schrenk12,Sharafedini23,Hamzehpour14,Fehr11,Sharafedini15,
Oliveira19,Hamzehpour14b}. The drawback of frequently adopting optimal paths is the 
increased probability of their failure by congestion, overload, etc. To explore the 
resilience of spatial networks, the optimal path crack (OPC) model was introduced 
\cite{Andrade09,Oliveira11,Fehr12} in which the site with the highest cost along 
the optimal path is blocked, the next emerging optimal path is identified, and 
this procedure is repeated until the entire network is disconnected. The roughness 
of the resulting backbone, i.e.\ the disconnecting path, has been found to lie in 
the same universality class for various physical processes with a fractal dimension 
of $d_f\,{\simeq}\,1.2$ and $2.5$ in two and three dimensions, respectively 
\cite{Andrade09,Oliveira11,Fehr12,Moreira12,Daryaei12}. 

\begin{SCfigure*}[\sidecaptionrelwidth][t!]
\centering
\includegraphics[width=1.35\linewidth]{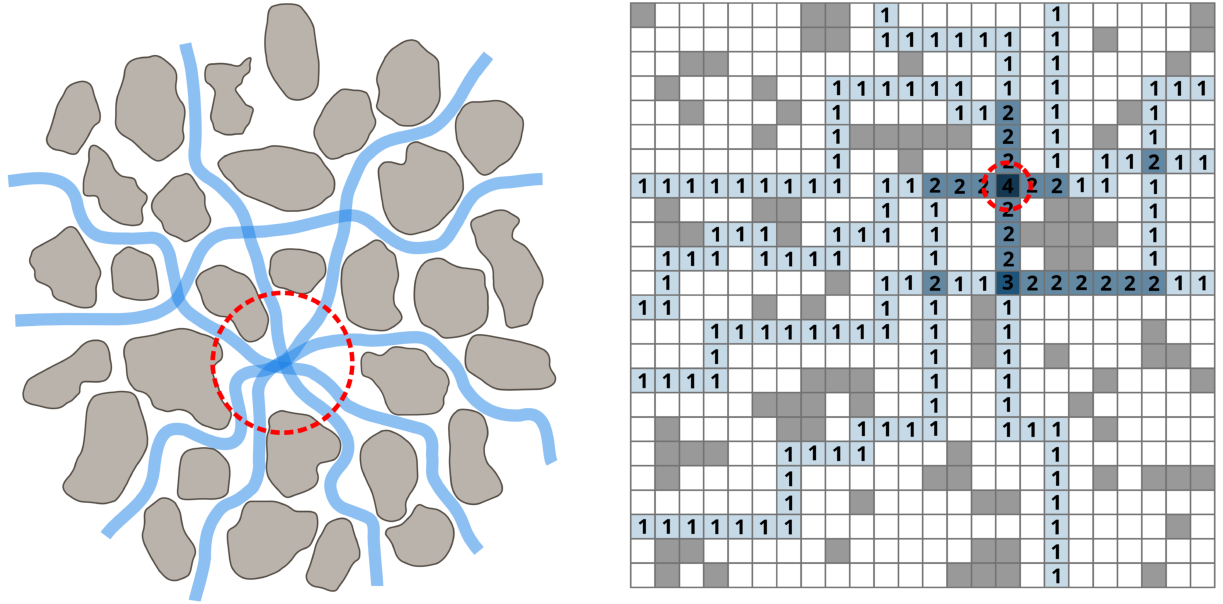}
\caption{(left) Schematic drawing of a porous medium. (right) Example of our porous 
lattice structures. A few optimal paths are shown in blue and the central transport 
hub is marked in red. The numbers in the right panel denote the nonzero centrality 
measures $\lambda(i,j)$ of the sites, defined in Eq.\,(\ref{Eq:Centrality}).}
\label{Fig:1}
\end{SCfigure*}

While the OPC model is based on processing the single optimal path information, past 
studies have shown that the relevance of network nodes for transport efficiency is 
better reflected in their betweenness centrality (BC) measure \cite{Newman01,Noh04,
Goh01,Kirkley18}, defined as the total number of passing through a node when the 
entire set of optimal paths between all possible source-sink pairs is considered. 
Optimization of transport by prioritizing the flow through nodes with a higher BC 
is accompanied by an enhanced vulnerability of the system to the cascading failure 
of transport hubs subject to stochastic blockage events or targeted attacks 
\cite{Albert00,Holme02,Bashan13,Buldyrev10}. It is expected that spatial networks 
display a higher degree of tolerance compared to, e.g., scale-free networks due 
to the possibility of nearly redundant routing between nodes. However, it is unclear 
how far a spatial network can adapt to successive blockage of transport hubs and 
what role the initial structure plays in approaching the onset of impenetrability.

Here we investigate the resilience of evolving spatial networks to effectively 
targeting the central transport hubs. By identifying the main hub in a porous 
lattice structure based on the BC measure of the nodes (see Fig.\,\ref{Fig:1}), 
we recursively block the hub and determine the next emerging one, until the 
system reaches the onset of impenetrability. We find a power-law scaling law 
for the required number of blocking steps with an exponent $d_f\,{\simeq}\,1.05$ 
which is outside the statistical error bar of the OPC universality class exponent. 
Our results reveal a universal dependence of the number of blocking steps on 
the initial occupation fraction of the network, which enables us to predict 
the behavior for arbitrary large networks and different initial network densities. 
Besides benefiting from the predictive power of our discovery of structural 
self-similarities, the computational costs can be further reduced since we show 
that our predictions are remarkably robust against partial structural information 
processing. Following the evolution of the shortest-path length statistics reveals 
a gradual crossover from random to correlated processes due to spatial exclusion 
effects. 
\smallskip\smallskip

\begin{SCfigure*}[\sidecaptionrelwidth][t!]
\centering
\includegraphics[width=1.15\linewidth]{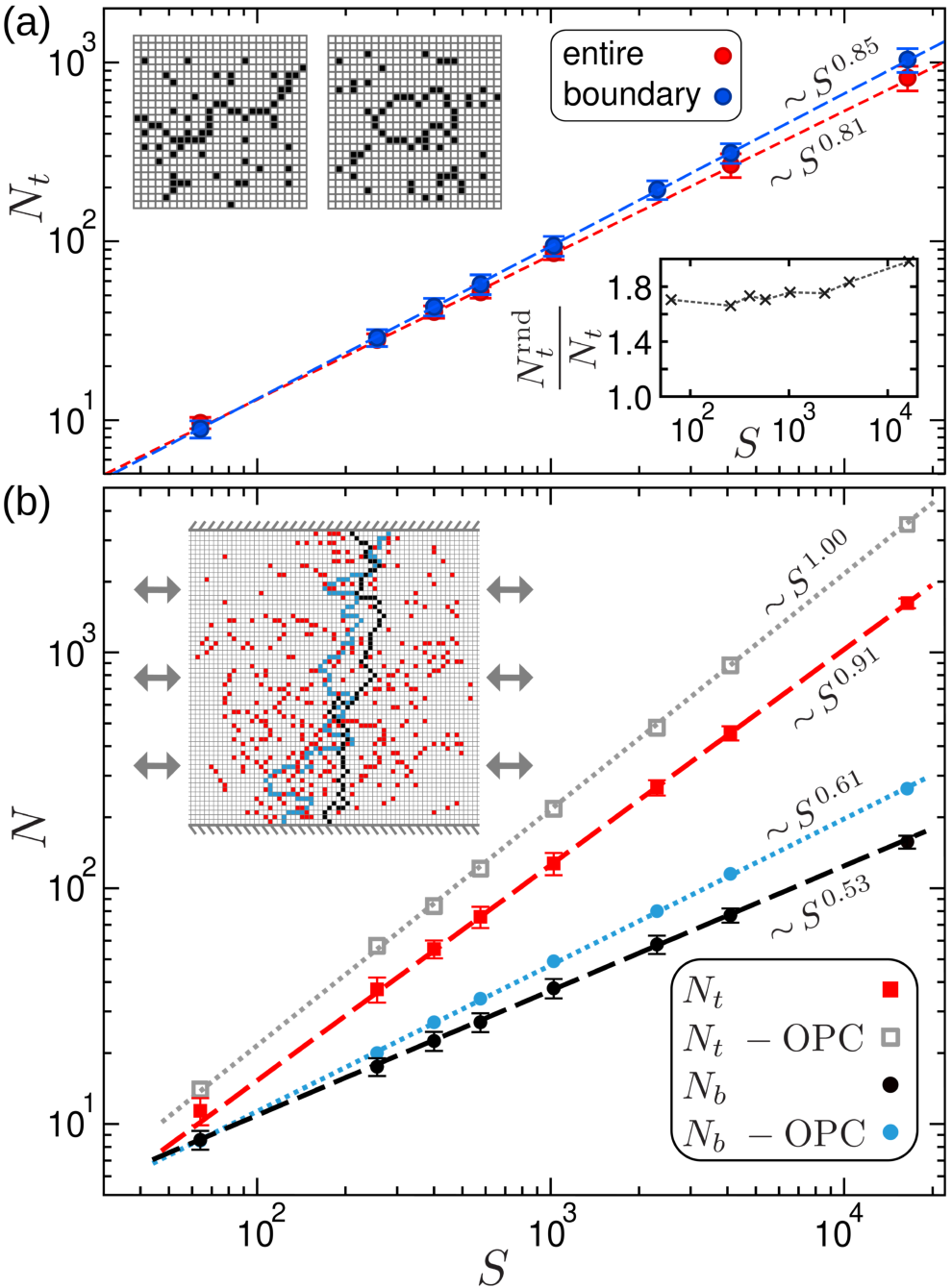}
\caption{(a) Total number of blocked sites $N_t$ vs the number of network nodes 
$S$ for $\phi_{_0}{=}\,0$ and source-sink pairs being on the boundaries (blue 
circles) or across the entire network (red circles). The lines are power-law 
fits to the data. Insets: (top) Examples of disconnected networks with loopless 
or looped backbones. (bottom) Ratio of $N_t$ for random and hierarchical site 
blockage versus $S$. (b) $N_t$ and backbone length $N_b$ vs $S$ for source-sink 
pairs being on the vertical boundaries. A comparison is made to the OPC model 
results. Inset: Typical realization of the network at the onset of impenetrability. 
The backbone, other blocked sites, and an example of the OPC backbone are shown 
in black, red, and blue respectively.}
\label{Fig:2}
\end{SCfigure*}

\begin{SCfigure*}[\sidecaptionrelwidth][b!]
\centering
\includegraphics[width=1.14\linewidth]{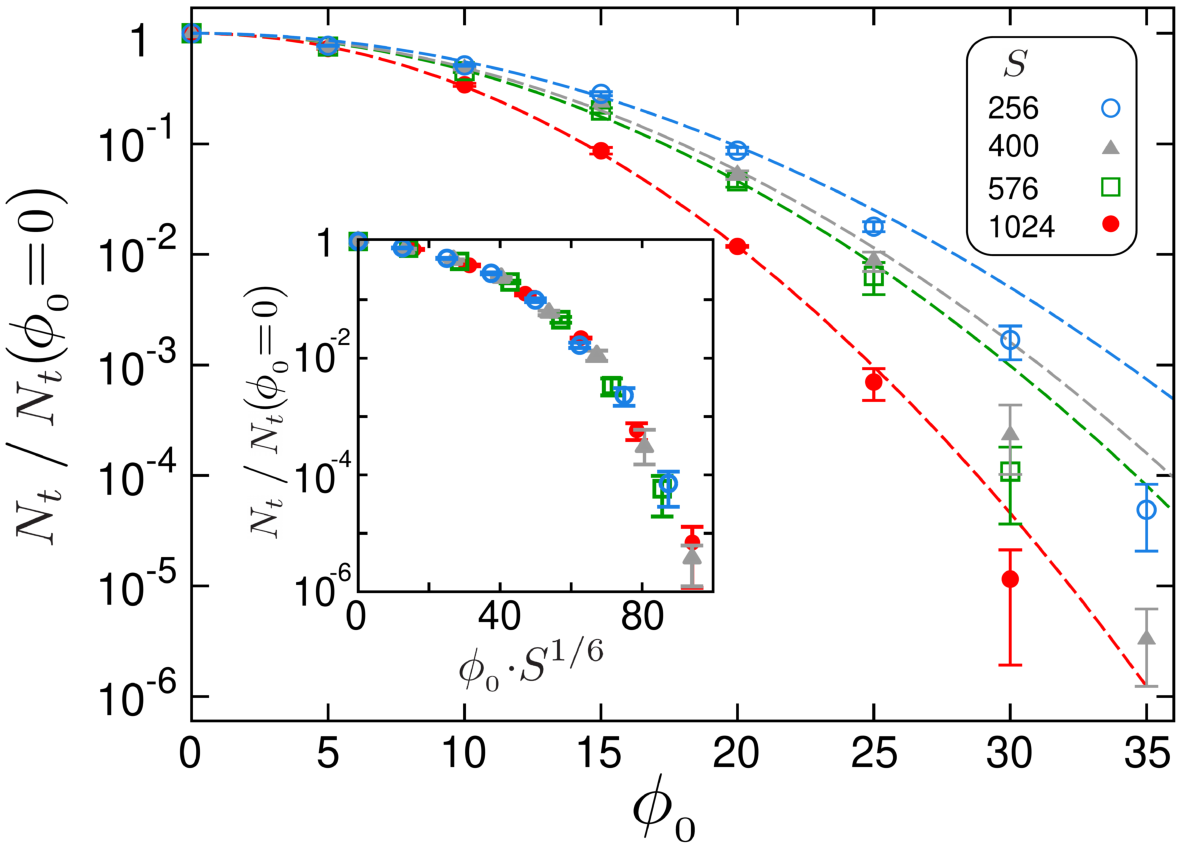}
\caption{Rescaled $N_t$ versus the initial occupation fraction $\phi_{_0}$ 
for different network sizes. The dashed lines represent overall Gaussian 
fits to the data. Inset: Data collapse of the number of blocking steps 
versus the rescaled initial occupation fraction for different network 
sizes.}
\label{Fig:3}
\end{SCfigure*}

\section{Porous lattice network model}

We consider a square lattice with $S$ sites and reflective boundary conditions 
in all directions. An initial fraction $\phi_{_0}$ of the sites are randomly 
occupied in such a way that the structure remains connected. The main transport 
hub of the network is determined by evaluating the optimal paths between every 
possible pair of the lattice sites, using the Dijkstra shortest-path algorithm 
\cite{Dijkstra59}. To avoid redundant shortest paths for each source-sink pair, 
a small random cost $c_{_{ij}}$ (${\ll}1{/}S$) is assigned to each empty site 
$(i,j)$ while $c_{_{ij}}{=}\,1$ for occupied sites. The shortest path between 
two nodes is the one among all possible paths that has the smallest sum over 
all $c_{_{ij}}$ costs of its sites. Next, we assign a centrality measure 
$\lambda(i,j)$ to each site, defined as
\begin{equation}
\lambda(i,j)= \displaystyle\sum_{k{\neq}l} \sigma\!_{_{kl}}(i,j), 
\label{Eq:Centrality}
\end{equation}
where $(k,l)$ indices run over all possible source-sink pairs. $\sigma\!_{_{kl}}(i,j)$ 
is $1$ if the shortest path between $k$ and $l$ nodes passes through the $(i,j)$ site 
and $0$ otherwise. The site with the highest $\lambda$ is identified as the main 
transport hub and turned into an occupied site by changing its $c_{_{ij}}$ to $1$. 
This procedure is then repeated until the network is disconnected. We define 
penetrability as having access from any existing source to any existing sink 
in the system. The blocking step at which this condition is violated is considered 
as the \emph{onset of impenetrability} in our simulations. The network is disconnected 
beyond the onset of impenetrability, meaning that there is at least one isolated source 
or sink in the system. We denote the total number of blocked sites until reaching the 
onset of impenetrability with $N_t$ and the length of the fault backbone (i.e.\ the line 
effectively breaking the network in two) with $N_b$. Although the exact computation of 
the centrality is extremely costly \cite{ErcseyRavasz10}, for the simulation results 
presented in this manuscript we have exhaustively determined the shortest path between 
all source-sink pairs (unless stated otherwise) to accurately identify the betweenness 
centrality measure for all nodes. Thus, the search for optimal path is repeated for 
$\mathcal{O}(S^2)$ pairs (i.e.\ linear lattice size to power four).
\smallskip\smallskip

\section{Results}

\subsection{Structural self-similarities}

We first consider a general case of an isotropic medium without a preferred flow 
direction and suppose that every pair of nodes across the entire network can be 
a possible source-sink transport set. Starting with an empty lattice ($\phi_{_0}
{=}\,0$), the number of blocking steps to reach the disconnection point is 
shown in Fig.\,\ref{Fig:2}(a) for different network sizes. Interestingly, 
$N_t$ scales as $N_t\,{\simeq}\,S^{\beta{/}2}$ with an exponent $\beta\,{=}
\,1.61\,{\pm}\,0.01$. In contrast to the fault backbone in the OPC models, 
here the backbone does not necessarily pass across the network; the isotropicity 
of the endpoint positions of optimal paths enhances the transport through 
the bulk, favoring the formation of looped backbones in the central regions 
[see inset of Fig.\,\ref{Fig:2}(a)]. 

By exclusively choosing the source-sink pairs on the lattice boundaries, 
material fluxes through the bulk of the system reduce. This suggests that 
the blocked sites are distributed more uniformly across the network in 
this case, which enhances the probability of loopless backbone formation 
and increases the total number of required blocking steps to disconnect 
the network. The results shown in Fig.\,\ref{Fig:2}(a) confirm this 
hypothesis: $N_t$ is larger for source-sink pairs on the boundaries 
compared to being across the entire network (similarly for the backbone 
length $N_b$; not shown). Also the fractal dimension slightly increases 
to $\beta\,{=}\,1.71{\pm}0.03$. In the lower inset of Fig.\,\ref{Fig:2}(a), 
a comparison is made to the inefficient strategy of randomly blocking 
the lattice sites; the inefficiency of this approach in disconnecting 
the structure slightly grows with increasing number of network nodes. 

To clarify whether the fractal dimension of the backbone based on our BC 
measure lies in the OPC universality class, we further limit the choices 
of source-sink pairs to the vertical boundaries as in the OPC models; 
see inset of Fig.\,\ref{Fig:2}(b). Such an asymmetric transport eliminates 
the chance of looped backbone formation, resulting in larger $N_t$ and 
$N_b$ values than those for the previously discussed symmetric conditions.
Figure\,\ref{Fig:2}(b) reveals that our BC-based blockage brings the 
network to the impenetrability limit with a remarkably less number 
of blocking steps $N_t$ and a lower power-law exponent ($\beta\,{=}\,
1.82\,{\pm}\,0.01$ vs $2.00\,{\pm}\,0.01$) compared to the OPC model.
Moreover, our backbones are visibly smoother than the OPC ones (see 
inset). The backbone length scales in both cases as $N_b\,{\simeq}\,
S^{\gamma{/}2}$ but with a distinctly lower exponent $\gamma\,{=}\,
1.06\,{\pm}\,0.01$ in our simulations compared to $1.22\,{\pm}\,0.02$ 
in the OPC models \cite{Andrade09,Oliveira11,Fehr12,Moreira12,Daryaei12}, 
demonstrating that the backbone roughnesses of the two models belong 
to different universality classes.

\begin{SCfigure*}[\sidecaptionrelwidth][t!]
\centering
\includegraphics[width=1.14\linewidth]{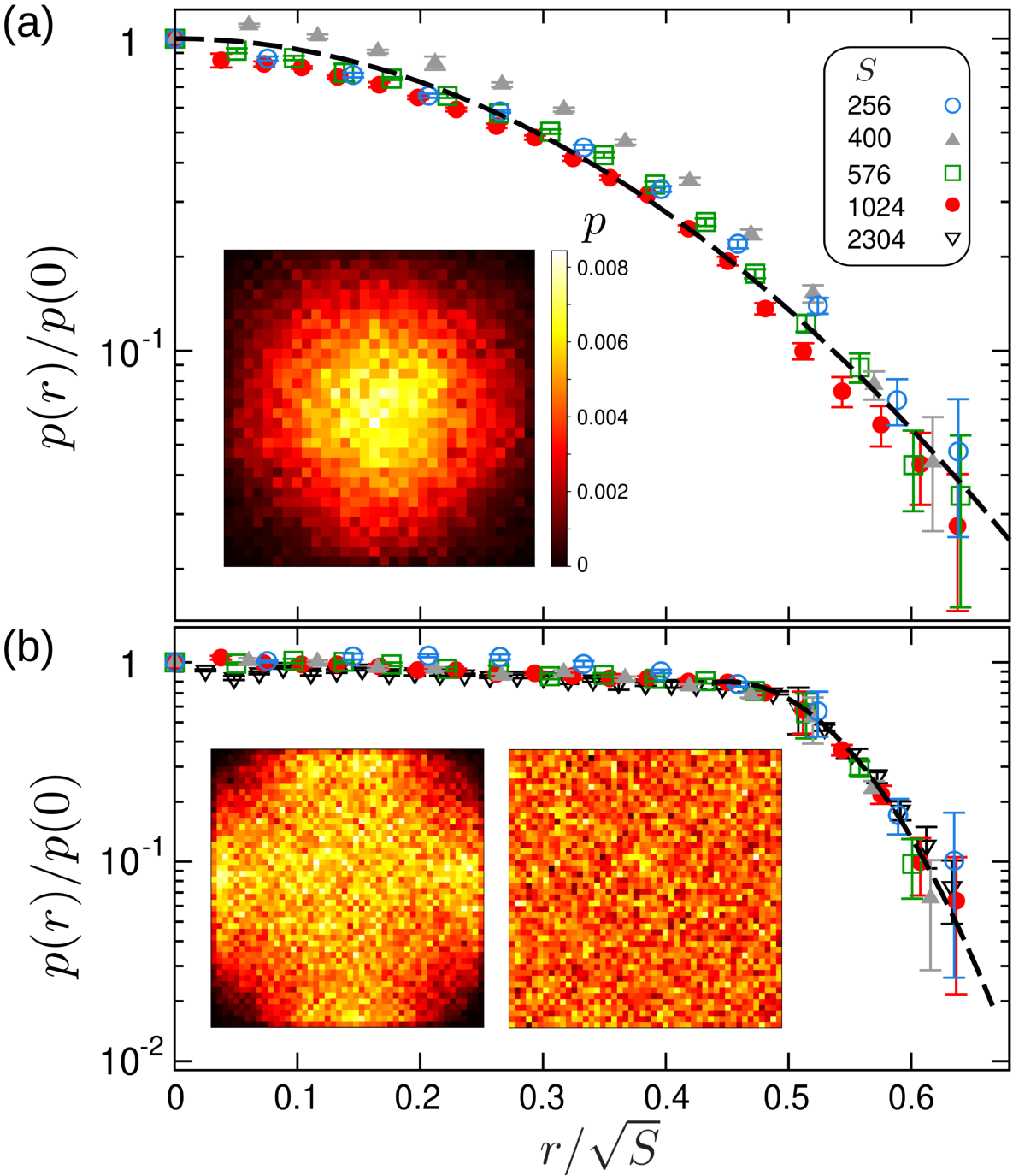}
\caption{Blocking probability $p(r)$ of sites versus their radial 
distance $r$ from the center of the structure for different network 
sizes and source-sink pairs being (a) across the entire network 
or (b) on the boundaries. The dashed lines show a Gaussian fit, 
Eq.\,(\ref{Eq:pr}). The insets show the spatial distribution of 
$p(r)$ (The right inset of panel (b) represents a random blockage 
process for comparison).}
\label{Fig:4}
\end{SCfigure*}

Next, we address the role of the initial occupation fraction $\phi_{_0}$ 
of the structure (i.e.\ starting with $N_{_0}{=}\,\phi_{_0}S$ randomly 
occupied sites) in approaching the impenetrability limit. Here we show 
the results for source-sink pairs being across the entire network, but 
similar conclusions can be drawn for other boundary conditions. For a 
purely random blockage process, the average number of blocking steps 
to disconnect the structure follows $N_t(\phi_{_0}\!)\,{=}\,N_t(\phi_{_0}
{=}0){-}\phi_{_0}S$, imposing an upper limit $\phi_{_0}^{\,\text{max}}
\,{=}\,N_t(\phi_{_0}{=}0){/}S$ for the initial occupation fraction of 
the network to remain connected. However, we can increase $\phi_{_0}$ 
beyond $\phi_{_0}^{\,\text{max}}$ by selecting only the connected 
initial structures. The BC-based hierarchical site elimination procedure 
is then applied to disconnect these structures. Figure\,\ref{Fig:3} 
displays a network size-dependent decay of $N_t$ versus $\phi_{_0}$, with 
a faster than a Gaussian tail (which can be roughly captured by a Cauchy 
distribution, probably reflecting the heterogeneity of the initial 
sampling). At a given $\phi_{_0}$, the deviations grow with decreasing 
$S$, along with the increasing difficulty in finding connected initial 
structures. Importantly, the data collapses onto a single master curve 
for different $S$ when $\phi_{_0}$ axis is rescaled by $S^{1{/}6}$; see 
inset of Fig.\,\ref{Fig:3}. This implies that the number of steps to 
block a network with a given initial occupation fraction $\phi_{_0}$ 
scales as
\begin{equation}
N_t(\phi_{_0}) = S^{\beta{/}2} \widetilde{N}_t(\phi_{_0}{\cdot}S^{1{/}6}), 
\label{Eq:NphiScaling}
\end{equation}
with $\widetilde{N}_t(0)\,{=}\,1$.
\smallskip\smallskip

\subsection{Evolution of spatial correlations}

\begin{figure}[t]
\centering
\includegraphics[width=0.47\textwidth]{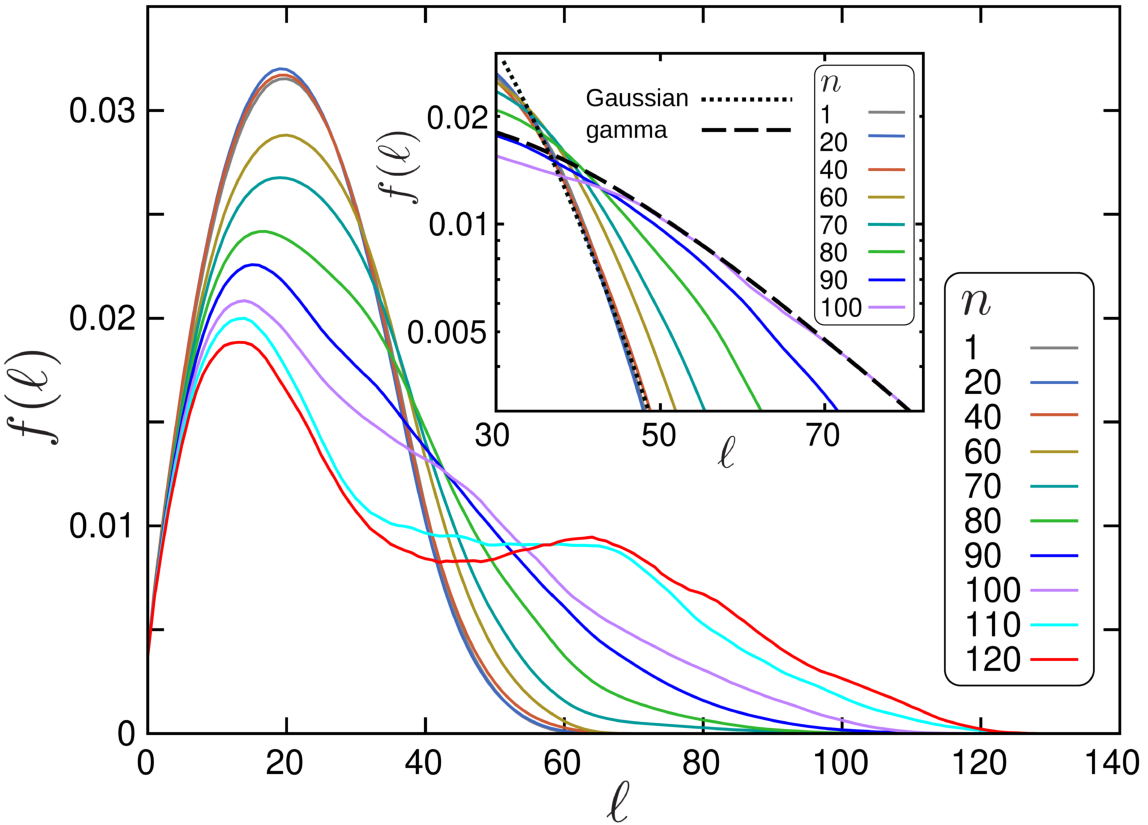}
\caption{Probability distribution of the shortest-path length $\ell$ at 
different stages $n$ of the blocking process. The inset zooms in on the 
tail behavior. The dotted and dashed lines indicate Gaussian and gamma 
distributions as a guide to the eye.}
\label{Fig:5}
\end{figure}

Past studies attempted to evaluate the BC distribution on various 
static spatial networks \cite{Kirkley18,Verbavatz22}, suggesting, e.g., a fast 
BC decay of the form $1{-}r^2{/}S$ ($r$ being the distance from the network 
center) in highly porous structures. However, it is unclear how the spatial 
distribution of centralities evolves during a cascading failure of high BC 
nodes under a targeted attack. By measuring the blocking probability $p(r)$ 
in our dynamically evolving porous lattice structures, we find that the 
behavior is well captured by a Gaussian radial decay  
\begin{equation}
\displaystyle p(r)\sim \text{exp}\Big[-\frac12 \big(\frac{r}{\sqrt{S}{/}2}
\big)^2\Big]
\label{Eq:pr}
\end{equation}
for source-sink pairs being across the entire network. Choosing the pairs 
on the boundaries diminishes the importance of the central regions, leading 
to a plateau at small $r$; nevertheless, the tail of $p(r)$ towards the 
corners still satisfactorily follows Eq.\,(\ref{Eq:pr}); see Fig.\,\ref{Fig:4} 
and insets. As a result of the rapid radial decay of the importance of sites 
for homogeneously distributed sources and sinks, blockage of the transport 
hubs mostly occurs in the central region which leads to looped backbone 
formation. By distributing the sources and sinks on the boundaries, the 
shortest paths do not necessarily pass through the center of the system 
and the importance of the sites becomes spatially more uniform, which 
promotes loopless backbones. To better understand the morphological 
transition from looped to loopless backbones (e.g.\ to clarify whether 
it is a continuous or discontinuous transition), the sinks and sources 
can be gradually pushed away from the center of the network \footnote{We 
thank an anonymous referee for the valuable suggestion.}. However, 
such a study is computationally costly and beyond the scope of the present 
work. A similar morphological transition has been reported in the context of 
optimal transportation networks \cite{Aldous19}. Compared to static spatial 
networks, the slower decay of $p(r)$ in our evolving structures in the course 
of blocking process indicates growing spatial correlations and constraints. A 
higher chance of site elimination in central regions creates bottlenecks and, 
thus, new high BC sites in these regions. This process self-amplifies and 
leads to a slower spreading of the emerging transport hubs across the network. 
Formation of bottlenecks in the bulk causes longer shortest-paths $\ell$, 
affecting the shortest-path length statistics. The shortest-path length 
distribution $f(\ell)$ is shown in Fig.\,\ref{Fig:5} at different stages 
$n$ of the blocking process. It can be seen that $f(\ell)$ broadens during 
the blocking process with a tail gradually evolving from a Gaussian towards 
a gamma distribution $\frac{\ell^{\sigma{-}1}}{\lambda^{\sigma}\Gamma(
\sigma)}\text{exp}(-\ell{/}\lambda)$ associated with random events in 
the presence of self-correlations \cite{Vliegenthart06,Shaebani17,
Deboeuf13} (see inset). $f(\ell)$ eventually develops a bimodal shape 
at the latest blocking stages before the onset of impenetrability 
because the network practically splits into two islands whose 
connection through the remaining bottlenecks creates the secondary 
peak of $f(\ell)$ at longer shortest-paths. 
\smallskip\smallskip

\subsection{Robustness of results}

\begin{figure}[t]
\centering
\includegraphics[width=0.47\textwidth]{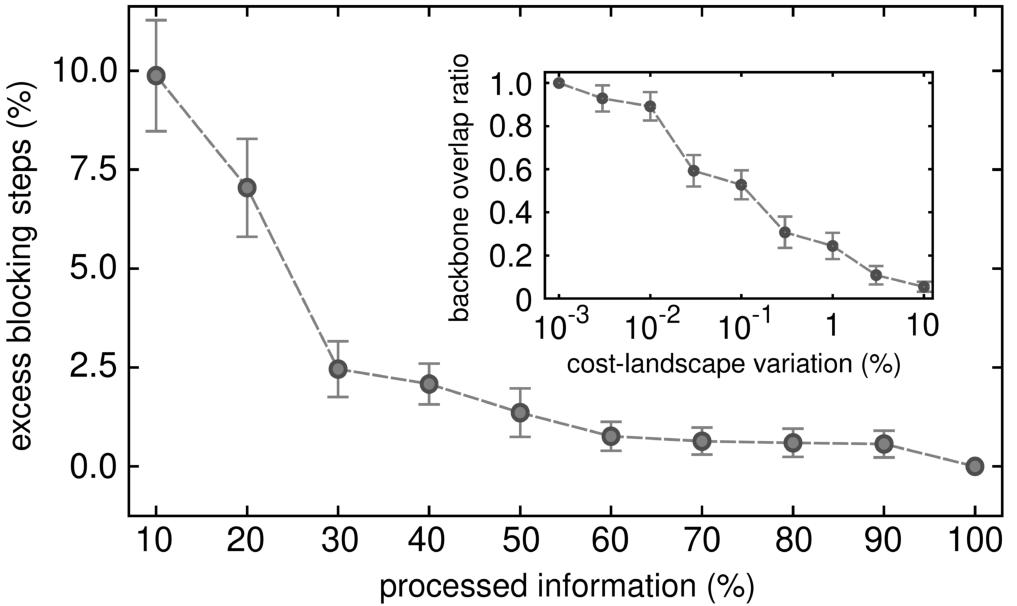}
\caption{Excess number of required blocking steps in terms of the fraction 
of randomly processed source-sink pairs to evaluate the centrality measures 
$\lambda(i,j)$ at each blocking step, for pairs being across the entire 
network with $S\,{=}\,4096$ nodes. Inset: Deviation of the backbone path 
from the original one upon increasing the variation range of the random 
costs $c_{_{ij}}$ of empty sites around their original values.}
\label{Fig:6}
\end{figure}

Finally, we address the stability of our findings against incomplete 
structural information. This includes uncertainty in the initial cost 
landscape and/or a partial processing of the structural information.
We first modify the initial conditions by changing the random cost 
$c_{_{ij}}$ of each empty site $(i,j)$ within a uniform variation range 
around the original value. Starting from this new cost landscape, we 
obtain the number of blocking steps and the backbone path and compare 
them with those of the original cost landscape. $N_t$ does not show 
systematic dependence on the variations (or inaccuracies in the 
measurements) of the cost landscape (not shown), indicating that 
our prediction of transport resilience and onset of impenetrability 
is reliable. However, the results shown in the inset of Fig.\,\ref{Fig:6} 
demonstrate that the backbone path is highly sensitive to the cost 
landscape variations; even $1\%$ change of $c_{_{ij}}$ values 
results in a new backbone with less than $25\%$ overlap with the 
original backbone. Thus, a precise knowledge of the cost landscape 
is crucial for predicting the exact backbone path in porous lattice 
networks. Figure\,\ref{Fig:6} also reveals that the results are 
highly robust against an incomplete processing of the structural 
information, i.e., if only a fraction of the source-sink pairs 
are randomly chosen to evaluate the centrality measures $\lambda(i,j)$ 
at each blocking step. We vary the fraction of processed source-sink 
pairs for a network with $S\,{=}\,4096$ nodes from $100\%$ (${\sim}\,
10^7$ pairs) to $10\%$ (${\sim}\,10^6$ randomly chosen pairs) to assess 
the robustness of the results. It can be seen that by processing of 
only $25\%$ of all possible routes, the change in the number of blocking 
steps still remains below $5\%$. This promises the possibility of 
significant reduction of computational costs in the BC-based evaluation 
of main transport hubs according to our proposed approach.
\smallskip\smallskip

\section{Conclusion}

In conclusion, we have shown that the optimization of transport through 
evolving porous lattice structures based on the centrality measure of 
the nodes has the drawback of reduced adaptability and resilience of the 
spatial network during a cascading failure of main transport hubs subject 
to stochastic blockage events or targeted attacks. Our findings of universal 
resilience of spatial networks allow for predicting the number of blocking 
steps to reach the onset of impenetrability for different network sizes 
and occupation densities. Transport on dynamic porous lattice structures 
is typical of many problems in this class. We expect that our findings 
and conclusions hold for a broad range of real-world spatial networks 
and welcome work in this area. The reliability of our predictions upon 
considerable reduction of the amount of structural information processing 
can be combined with multiscale analysis approaches to further reduce 
the computational costs, which is crucial for practical and technological 
applications. The differences between the universality class of centrality-based 
node elimination and the OPC models deserve to be further studied on 
other topologies in spatial networks and for correlated cost landscapes. 
These remain exciting open challenges toward better understanding and 
designing of transport processes under disturbances and to improve the 
adaptation and recovery of transport capacity in dynamically evolving 
structures. \\

\noindent\textbf{Acknowledgments} \\

\noindent This work was supported by the Deutsche Forschungsgemeinschaft (DFG) 
via grants INST 256/539-1, which funded the computing resources at Saarland 
University. 

\bibliography{Refs-OptimalPaths}

\end{document}